\documentclass[aps,reprint,twocolumn]{revtex4-2}

\usepackage{graphicx}% Include figure files
\usepackage{dcolumn}% Align table columns on decimal point
\usepackage{bm}% bold math
\usepackage{color}
\usepackage{amsmath}
\usepackage{amssymb}
\usepackage{graphicx}
\usepackage[unicode=true,pdfusetitle,
 bookmarks=true,bookmarksnumbered=false,bookmarksopen=false,
 breaklinks=false,pdfborder={0 0 0},pdfborderstyle={},backref=false,colorlinks=true]
 {hyperref}
\hypersetup{
 pdfborderstyle={},citecolor=blue}

\usepackage[utf8]{inputenc}
\usepackage[T1]{fontenc}
\usepackage{mathptmx}

\begin{document}
\title{
Pulse instabilities in harmonic active mode-locking: \\ a time-delayed approach}

\author{Elias R. Koch}%
% \email{}
\affiliation{Institute for Theoretical Physics, University of M\"unster, Wilhelm-Klemm-Str. 9, 48149 M\"unster, Germany}
\author{Svetlana V. Gurevich}
\affiliation{Institute for Theoretical Physics, University of M\"unster, Wilhelm-Klemm-Str. 9, 48149 M\"unster, Germany}
\affiliation{Center for Nonlinear Science (CeNoS), University of M\"unster, Corrensstrasse 2, 48149 M\"unster, Germany}
\author{Julien Javaloyes}
\affiliation{Departament de Física \& IAC-3, Universitat de les Illes Balears, C/ Valldemossa km 7.5, 07122 Mallorca, Spain} 
\begin{abstract}
We propose a time-delayed model for the study of active mode-locking that is valid for large values of the round-trip gain and losses. It allows us to access the typical regimes encountered in semiconductor lasers and to perform an extended bifurcation analysis. Close to the harmonic resonances and to the lasing threshold, we recover the Hermite-Gauss solutions. However, the presence of the linewidth enhancement factor induces complex regimes in which even the fundamental solution becomes unstable. Finally, we discover a global bifurcation scenario in which a single pulse can jump, over a slow time scale, between the different minima of the modulation potential.
\end{abstract}

\maketitle
\begin{figure}[t!]
	\centering
	\includegraphics[width=1\columnwidth]{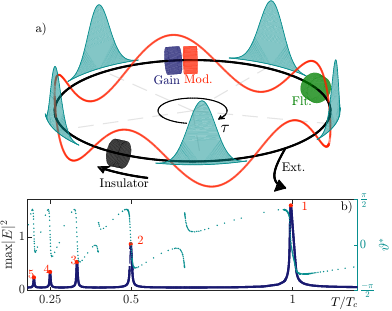}
	\caption{a) A schematic of an AML ring laser containing five pulses, i.e. $\tau\approx 5T$.
	b) Resonance peaks (dark blue) with relative positions (turquoise) of the pulses to potential minimum. The field intensity maximum is shown as a function of the ratio of the driving period $T$ and the first cavity resonance $T_c$. Parameters are: $(\gamma,\,\kappa,\,\alpha,\,G_0,\,\Gamma,\,m,\,\psi,\,\tau)=(40,\,0.4,\,0,\,1,\,1,\,0.15,\,0,\,1)$. }\label{fig:1}
\end{figure}

Manipulating and shaping light is of great importance for many applications and mode-locked (ML) lasers are widely employed as high-coherence light sources~\cite{ST_OLT_15} utilized for generating ultrashort light pulses and frequency combs~\cite{RHWC_NC_16,YZC_OE_20} across various applications 
ranging from medicine to laser metrology~\cite{WSY_APLP_16,HMCD_JLT_00,HZU_IEE_01,HUH_PRL_00,UHH_N_02,K_Nt_03}. A ML state is defined by the coherent superposition of many lasing modes in a cavity. When these modes adopt a specific phase relation, their coherent superposition may result in a train of short optical pulses whose repetition period corresponds to the time of flight in the cavity. However, such regimes are not obtained without implementing a specific mechanism favoring pulsed emission. Active mode-locking (AML) involves either an electro-optic or an acousto-optic intra-cavity modulator~\cite{TK_IEE_09, CDDC_OL_09,KKL_PhRes_17,KKWK_OL_94,PYZ_OC_02,ZHM_OC_05}, see Fig.~\ref{fig:1}~(a). 
If the modulation frequency is resonant with the separation between longitudinal modes, sidebands are created causing modal interactions that eventually lead to the emission of a pulse train. The main advantage of AML is its electrical tunability and the possibility to control the output pulse shape, energy and repetition rate~\cite{KS1_JQE_70,BWX_AFM_18}. 
However, intensity modulators incur high costs and increased complexity. Notably, a recently presented graphene-based electro-optic modulator offers improved performance and cost efficiency~\cite{BWX_AFM_18}. Further, the repetition rate in AML is limited by the highest frequency at which the modulator can be driven, a few tens of GHz, and pulses cannot be as short as in passive mode-locking (PML) using a saturable absorber~\cite{H-JSTQE-00}. Repetition-rate multiplication~\cite{SOH_APL_98} and pulse shortening \cite{TK_IEE_09,KKK_JOptSocAmB_95} can mitigate this problem.
% through additional effects. %and filters.
% 

The Haus master equation (HME) is a widely used theoretical model for both AML and PML that is derived from general principles using the assumptions of small gain, losses, as well as weak spectral filtering~\cite{H-JSTQE-00}. In particular, in the case of AML, it consists in restricting the analysis of the field to a small temporal interval around the pulse whose evolution, round-trip after round-trip, is described by a single partial differential equation (PDE). The latter is then coupled to the equations governing carrier dynamics. The HME was originally derived for gain media that are slowly evolving on the timescale of the cavity round-trip, resulting in a quasi-uniform gain temporal profile within the cavity. However, recent generalizations of the HME~\cite{PGG_NC_20,HLGJ_OL_20,NV_PRE_21} preserve carrier memory from one round-trip towards the next, allowing for the description of complex pulse trains such as the Q-switched transitions or a train of several equidistant pulses in the cavity, the so-called harmonic ML (HML). However, when applied to a particular design, the HME often provides only qualitative predictions due to the many simplifying hypotheses involved and, in general, the question of how to derive the HME for a specific laser design still remains open. On another hand, the models derived from first principles and based on time-delayed systems (TDSs)~\cite{VTK-OL-04,VT_PRA_05,MB-JQE-05} not only naturally allow to analyze scenarios with large gain and losses per round-trip typical of semiconductor lasers but they are also an essential tool to unveil connections between the pulsed regime and all the other solutions. TDS models are extensively applied to describe pulse dynamics in e.g., ring PML lasers~\cite{VT_PRA_05}, Fourier domain ML lasers~\cite{SKO-OE-13}, various PML semiconductor lasers with complex design~\cite{MB-JQE-05,MJB-PRL-14,MJB-JSTQE-15,Avrutin2019,SCM-PRL-19,BSV-OL-21,SHJ-PRAp-20,HMJGL-PRAppl-20} and photonic crystal ML lasers~\cite{HBM-OE-10}.

In this letter, we propose a time-delayed description for AML with intracavity loss or phase modulation which allows for the efficient study of the AML dynamics including gain dynamics and multipulse regimes as well as extensive numerical simulations and bifurcation analysis.
A sketch of the system is presented in Fig.~\ref{fig:1}~(a). It consists in a ring cavity with round-trip time $\tau$ that contains a gain section, a phase or an intensity modulator and a linear bandpass filter. The latter models, beyond the intentional use of an element to control pulse width, the etalon effect created by parasitic reflectivities within intracavity elements. It can also phenomenologically represents the finite bandwidth of the gain element. This approach was used e.g., in \cite{KZM-PRL-99} and can be derived rigorously, see \cite{PGG_NC_20}. While it was shown in \cite{KS2_JQE_70} that a small tilt of the modulator can mitigate interferometric effects and improve pulse width, this strategy is not always accessible, e.g. in fiber coupled systems \cite{PGG_NC_20}.
\begin{figure}[t!]
	\centering
	\includegraphics[width=1\columnwidth]{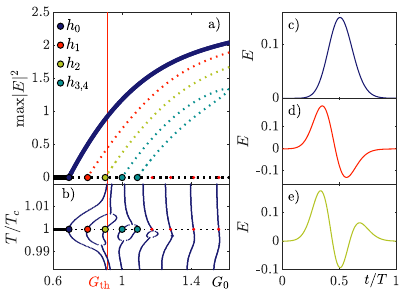}
	\caption{
a) First weakly nonlinear HGMs $h_l$ of the TDS~(\ref{eq:AML1},\ref{eq:AML2}) emerging from the off solution. Only the $h_0$ mode is stable. b) Dynamics of the bifurcation points of the off solution in the $(G_0,T/T_c)$ plane. The higher order modes emerge in pairs. Around resonance, the onset of pulses is notable below $G_{th}$. (c-e) Profiles of the modes $h_{0,1,2}$. Other parameters as in Fig.~\ref{fig:1}.}
	\label{fig:2}
\end{figure} 
Finally, an optical isolator ensures unidirectional propagation.
The derivation of our model follows the lines of ~\cite{VT_PRA_05}, cf. Supplement I. 
There, we linearize the gain and the carrier recombination as a function of the population inversion $G$
and assume that the filter bandwidth is much smaller than the gain broadening.
The evolution equations for the field $E$ at the filter output and $G$ read
\begin{align}
\frac{\dot{E}}{\gamma}+E&=\sqrt{\kappa(t)}e^{\left(1-i\alpha\right)G\left(t-\tau\right)/2+i\psi}E\left(t-\tau\right)\,, \label{eq:AML1}\\
\dot{G}&=\Gamma\left(G_{0}-G\right)-\left(e^{G}-1\right)\left|E\right|^{2}\label{eq:AML2}.
\end{align}
Here, $G_0,\Gamma,\alpha,\gamma$ are the pumping and gain recovery rates, the linewidth enhancement factor, the bandwidth of the spectral filter, respectively. 
The evolution of the round-trip phase $\psi$ could model small cavity length fluctuations due to e.g. thermal or mechanical noise. The fraction of light intensity kept in the cavity at the output coupler is $T_o$ while the intensity modulator transmission function is $T_i(t)$, leading to $\kappa(t)=T_o T_i(t)$. 
While our approach is not limited in the shape nor in the amplitude of the intensity modulation, we asssume here that  
 $T_i(t)=\overline{T}+\delta T \cos\left(\omega_m t\right)$ with $\omega_m$ the modulation frequency.
For small values of $\delta T/\overline{T}$ we observe that $\sqrt{\kappa(t)}=\sqrt{\kappa_0}\left[1+m\cos\left(\omega_m t\right)\right]$. It defines the modulation depth $m=\delta T/(2\overline{T})$ and the average transmission $\kappa_0 = T_o \overline{T} $. 
We note that a phase modulation is obtained setting $\sqrt{\kappa(t)}=\sqrt{\kappa_0}\exp\left[i\delta \phi \cos\left(\omega_m t\right)\right]$.
The parameters in Fig.~\ref{fig:1} correspond to a $30\,$cm cavity, a carrier lifetime of $1\,$ns and a filter bandwidth of $25\,$GHz. 
Note that Eqs.~(\ref{eq:AML1},\ref{eq:AML2}) can easily be adapted to other ring geometries with fast gain dynamics like quantum cascade lasers~\cite{PSK-Nature-20} by adiabatically eliminating the gain and, while in this letter we focus on semiconductor lasers, setting larger values of ($\tau,\Gamma^{-1}$) and $\alpha=0$ would correspond to e.g., Erbium fiber lasers.

For $m\neq 0$, the modulator creates a potential with $n$ slots, where
$n=\tau/T$ with $T=2\pi/\omega_m$, cf. Fig.~\ref{fig:1} for $n=5$. Note that the TDS~(\ref{eq:AML1},\ref{eq:AML2}) possesses the intrinsic resonance frequency $\omega_c=\frac{2\pi}{\tau}\frac{\gamma}{1+\gamma}$ (cf. Supplement I). The respective cavity resonance period $T_c=\frac{2\pi}{\omega_c}$ is slightly larger than the delay $\tau$ due to finite interaction times. For any choice $\omega_m\neq\omega_c$ one thus obtains asynchronous modulation and a resonance scan can be conducted by varying $\omega_m$ while keeping $\tau$ constant. Previous works such as \cite{KZM-PRL-99} employed a walk-off drift in a HME to model this effect.
The occurrence of resonance peaks for $n\in[1,5]$ is depicted in Fig.~\ref{fig:1}~(b). Following the resonance curve, the value of $\vartheta^*=\omega_m t^*$ indicates the pulse position $t^*$ relative to the transmission maximum. As the resonance is crossed, the pulse passes from being in advance to being delayed with respect to the latter. Between resonances the delayed pulse vanishes, being replaced by a newly arising advanced pulse. 
\begin{figure}[t!]
	\centering
	\includegraphics[width=1\columnwidth]{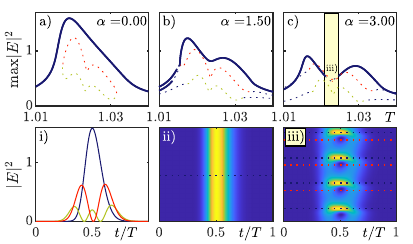}
	\caption{(a-c) Resonance curves of $h_0$ (blue), $h_1$ (red) and $h_2$ (green) modes, shown in i), for different values of $\alpha$. The curves reconnect and in c) a small region emerges, where both modes are unstable leading to oscillatory states between different modes. (ii-iii) Excerpt of the last 400 from $2.0\times10^5$ round-trips of a long time simulation for
	the $h_0$ mode and an oscillatory state between $h_0$ and $h_1$, indicated by colored dotted lines, respectively. Other parameters as in Fig.~\ref{fig:1}.
}
% \includegraphics[width=1\columnwidth]{fig3new.pdf}
% JJ:HOW MANY ROUNDTRIPS IN FIG3 II) AND III)
% EK: 400, but it is only the final roundtrips of a longer simulation
	\label{fig:3}
\end{figure} 
First, we choose $\alpha=0$ and consider a single pulse in the cavity, i.e. $T\approx\tau$. The TDS ~(\ref{eq:AML1},\ref{eq:AML2}) correctly predicts the emergence of the family of weakly nonlinear Hermite-Gauss modes (HGMs) $h_l$~\cite{HP-JQE-68, KS1_JQE_70}, that stem directly from the off solution $(E,\,G)=(0,\,G_0)$, cf. Fig.~\ref{fig:2}~(a). Using the path-continuation~\cite{DDEBT}, we followed the first five HGMs $h_0-h_4$ as a function of the pumping rate $G_0$. 
We implemented the non-autonomous TDS ~(\ref{eq:AML1},\ref{eq:AML2}) by setting $\dot{t}=1$.  
We notice that only the fundamental Gaussian mode $h_0$ is stable (solid blue line). As expected, due to the presence of the modulation, the lasing threshold for the $h_0$ mode is below~\cite{KS1_JQE_70} the value $G_{th}=-\log(\kappa_0)\simeq 0.92$ (cf. red line). Our reference point $G_{th}$ corresponds to the average transmission of the modulator and not its maximal value like e.g. in \cite{KZM-PRL-99}. The mode threshold gain is slightly above CW threshold for maximal transmission.

We depict in Fig.~\ref{fig:2}~(b), how the bifurcation points emerging on the off solution evolve with $G_0$, if the modulation period $T$ is varied. 
The bifurcation points (cf. colored circles in a)) occur along the horizontal black line at $T/\tau=1.025$. When $T$ is changed, the bifurcation point for the $h_0$ mode approaches $G_{th}$ as one leaves the resonance. There, the $h_0$ profile as depicted in panel c) deforms into a simple harmonic solution while the higher order modes disappear pairwise outside the resonance. Note that the resonance maxima of the different HGMs are generally not equal and also vary with $G_0$. The corresponding mode profiles get increasingly asymmetric far from the resonance (cf. Fig.~\ref{fig:2}~(e)). At threshold, the modes already experience an asymmetric potential if $\vartheta^*\neq 0$ while further asymmetry is expected due to gain dynamics for higher pumping \cite{JCM-PRL-16,PGG_NC_20}.

%%%%-------------------HME---------------------------------------------
To gain a better understanding of the emergence of the HGMs and to connect our analysis with former works such as~\cite{HP-JQE-68,KS1_JQE_70}, we derived a HME using the lasing threshold as an expansion point (cf. Supplement I). For $\omega_m=\omega_c$, the obtained HME correctly predicts the emergence of a family of HGMs. Their branching points over the off solution occur at the gain values $G_{th}^{(l)}=G_{th}-2m+\frac{\sqrt{m}\omega_m}{\gamma}(2l+1)$. At resonance, the pulses are located at the minimum of the potential which is the maximum of the modulation. Ergo, the onset for AML is located below the unmodulated laser threshold $G_{th}$ by a factor proportional to the modulation depth $m$. We evaluated $G_{th}^{(l)}$ for the first three modes and obtained $G_{th}^{(0,1,2)}=(0.676,0.794,0.913)$. This is around $1\%$ error in comparison with the respective bifurcation points in Fig.~\ref{fig:2} that are located at $G_0=(0.692,0.800,0.902)$.

%------ recombination of modes as a function of alpha
A more complex scenario is encountered for non-vanishing values of $\alpha$. We note in Fig.~\ref{fig:2}~(a,b) that the modes $h_l$ for $l\geq1$ are pairwise connected. A similar recombination of modes can be observed for higher $\alpha$, see Fig.~\ref{fig:3}. Here, the resonance curves for $h_0$ (blue), $h_1$ (red) and $h_2$ (green) are calculated for three different $\alpha$ values, cf. panels (a-c) and (i) for the modes profiles, whereas panel (ii) shows a zoom into the temporal evolution of the $h_0$. For growing $\alpha$, the maximum intensity of the $h_0$ decreases and the curve starts to develop a second maximum (cf. panels (a,b)). In addition, the distance between the $h_0$ and $h_1$ curve decreases such that they finally reconnect, see Fig.~\ref{fig:3}~(b). Further increase in $\alpha$ leads to another reconnection of the branches in the minimum, cf. Fig.~\ref{fig:3}~(c). Here, neither mode is stable. Long time simulations of at least $2\times10^5$ round-trips indicate that in this region complex states such as the oscillation between $h_0$ and $h_1$ occur, cf. Fig.~\ref{fig:3}~(iii). The same scan in $G_0$ for $\alpha=1.5$ (cf. Supplement I), indicates that $h_0$ loses stability at higher $G_0$. Further, the higher order modes are not pairwise connected any more. This is in agreement with Fig.~\ref{fig:3}, where the closed loops of the higher order modes split up at higher $\alpha$.
\begin{figure}[t!]
	\centering
	\includegraphics[width=1\columnwidth]{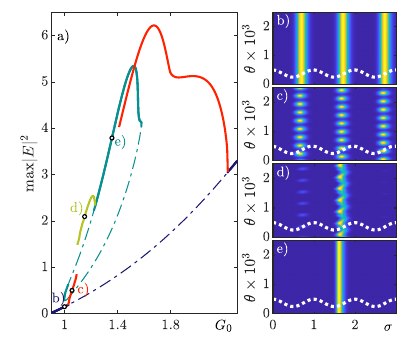}
	\caption{a) Bifurcation diagram of Eqs.~(\ref{eq:AML1},\ref{eq:AML2}) for $T=1.004$, $\Gamma=0.4$ and $\tau=3$ in the $(G_0,\,\textrm{max}(|E|^2))$ plane. Solid (dash-dotted) lines denote stable (unstable) solutions, respectively. Different colors correspond to the excerpt from the time trace of at least $2\tau\times10^5$ round-trips of b) HML$_3$ solution (blue), c) modulated HML$_3$ solution (red, see Visualization 1), d) Modulated single-pulse solution (green, see Visualization 2), e) single-pulse solution (turquoise). Parameters as in Fig.~\ref{fig:1}.
}
	\label{fig:4}
\end{figure} 
Now we consider a regime where $\tau\approx3T$, generating a train of three equidistant pulses, the so-called harmonic ML regime (HML$_3$), see Fig.~\ref{fig:4}~(b). The TDS~(\ref{eq:AML1},\ref{eq:AML2}) was solved numerically and the resulting time trace is plotted using the two-time representation where the fast timescale governs the dynamics within one round-trip, whereas the slow scale describes the dynamics from one round-trip to the next. Next, we followed the obtained HML$_3$ solution in the parameter space using path-continuation. Figure~\ref{fig:4}~(a) shows the resulting bifurcation diagram, where the maximum intensity is depicted as a function of $G_0$. There, the HML$_3$ solution (solid blue) becomes unstable (dotted-dashed blue) around $G_0=1$ in a torus bifurcation leading to a modulated HML$_3$ solution (solid red), see Fig.~\ref{fig:4}~(c) and Visualization 1. We followed the stable part of the branch using simulations of $2\tau\times10^5$ round-trips; for some $G_0$, the branch loses stability and  approaches the next stable attractor (yellow): a modulated single pulse solution in the three-slots potential, see Fig.~\ref{fig:4}~(d) and Visualization 2. Note that the single pulse has a small satellite located in the left next potential slot. This solution emerges from a single-pulse solution (turquoise) in a supercritical torus bifurcation, cf. Fig.~\ref{fig:4}~(e). This solution forms a closed loop in the parameter space and does not seem to be connected with other branches. Further increasing $G_0$, we observe that the single pulse solution coexists with the modulated HML$_3$, that is connected again to the stable HML$_3$ for larger $G_0$. Note that several of the observed solutions were reported in ~\cite{TK_IEE_09}, considering the HME with saturable gain. For lower $G_0$, the modulated HML$_3$ seems to terminate close to the single pulse branch. The absence of any bifurcation point on the latter indicates that, if this connection exists, it is of a global nature. Further, to get the first insights into the impact of $\alpha$ on the observed dynamical regimes, we conducted simulations for different values of $G_0$ and $\alpha$. The results are summarized in Fig.S4 of the Supplement I. There, one observes that the variety of dynamical solutions holds for higher $\alpha$ values.

\begin{figure}[t!]
	\centering
	\includegraphics[width=1\columnwidth]{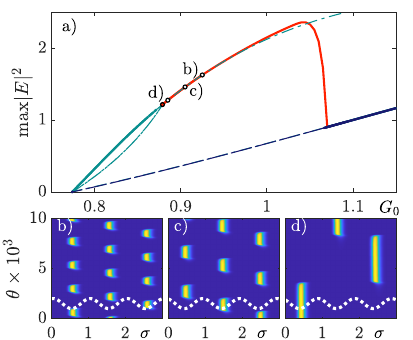}
	\caption{a) Bifurcation diagram of Eqs.~(\ref{eq:AML1},\ref{eq:AML2}) for $T=1.006$, $\tau=3$ and $\Gamma=0.5$ in the $(G_0,\,\textrm{max}(|E|^2))$ plane.
	HML$_3$ (dark blue), modulated HML$_3$ (green) as well as single-pulse solutions (turquoise) are shown. (b-d) Three time traces in two-time representation of the modulated HML$_3$ close to the global bifurcation (red point), see Visualization 3.
}
	\label{fig:5}
\end{figure}

Next, to study possible global transition, we slightly change the parameters and repeat the analysis for the HML$_3$ solution, see Fig.~\ref{fig:5}.
Here, we focus on the range of $G_0$, where only modulated HML$_3$ (red) and single-pulse solutions (turquoise) are stable, whereas the HML$_3$ is unstable (dash-dotted blue). In Fig.~\ref{fig:5}~(b-d) three time traces of $10^4$ from $2\tau\times10^5$ round-trips are presented. One observes a large increase of the modulation period as the bifurcation point (red point) is approached, indicating the global nature of this transition, cf. Visualization 3. The scaling of the oscillation period close to the critical $G_0$ value (cf. Supplement I) indicates a saddle-node infinite period bifurcation. A similar analysis for the parameter set of Fig.~\ref{fig:4} yields the characteristic scaling of a homoclinic bifurcation.

% %%%%-------------------HME---------------------------------------------
% Note, that to gain a better understanding of the emergence of the HG modes and to connect our analysis with former works such as ~\cite{KS1_JQE_70}, we derived a HME in the supplementary material (cf. Supplement I).

% --------conclusion

In conclusion, we proposed a TDS model for AML allowing for the study of the high gain and losses regimes typical for semiconductor lasers. This approach allowed us, by using a combination of path continuation and numerical simulations, to provide a detailed bifurcation analysis of how pulses emerge and evolve in AML, a topic that received comparatively less attention from the point of view of modeling. We showed that our AML model does not only correctly predict the emergence of basic solutions such as the HGMs, connecting our work with previous results obtained with the HME, but also show a complex scenario of modal interaction mediated by the linewidth enhancement factor and gain dynamics. Finally, we elucidated the transitions between different multipulse dynamical regimes, identifying two different global bifurcation scenarios. The latter could be relevant for AML lasers operating at high harmonic number.

$\,$

\small
%\textbf{Funding} 
%$\,$

\textbf{Disclosures} The authors declare no conflicts of interest.
$\,$

\textbf{Data availability} 
Data underlying the results presented in this paper are available in zenodo at \href{http://doi.org/10.5281/zenodo.10473547}{http://doi.org/10.5281/zenodo.10473547}. %Ref.~\cite{KGJ-OLZenodo-24}.
%Data underlying the results presented in this paper are not publicly 
%available  at this time but may be obtained from the authors upon reasonable request.
$\,$

\textbf{Supplemental document}  See Supplement I as well as Visualizations 1-3 for supporting content.
$\,$
%% add bibliography
%\bibliography{full_140723,AML_bib}
%\bibliographyfullrefs{full_140723,AML_bib}

\end{document}